\documentclass[runningheads]{llncs}
\usepackage{graphicx}
\usepackage{hyperref}
\begin{document}
\title{PARSE challenge 2022: Pulmonary Arteries Segmentation using Swin U-Net Transformer(Swin UNETR) and U-Net}
\titlerunning{PARSE Challenge 2022}
\author{Akansh Maurya\inst{1} \and
Kunal Dashrath Patil\inst{2} \and
Rohan Padhy\inst{2}\ \and
Kalluri Ramakrishna\inst{2}\ \and
Ganapathy Krishnamurthi\inst{2}
}

\authorrunning{Akansh Maurya et al.}
\institute{Robert Bosch Center of Data Science and AI, IIT Madras, India \email{akanshmaurya@gmail.com}\and
Indian Institute of Technology, Madras, India
\email{\{ed21s007,ed21s001, ed19d752, gankrish\}@smail.iitm.ac.in}}

\maketitle              % typeset the header of the contribution
\begin{abstract}
In this work, we present our proposed method to segment the pulmonary arteries from the CT scans using Swin UNETR and U-Net based deep neural network architecture. Six models, three models based on Swin UNETR, and three models based on 3D U-net with residual units were ensemble using a weighted average to make the final segmentation masks. Our team achieved a multi-level dice score of 84.36 percent through this method. The code of our work is available on the following link: \href{https://github.com/akansh12/parse2022}{https://github.com/akansh12/parse2022}. This work is part of the MICCAI PARSE 2022 challenge. 

\end{abstract}
\section{Introduction}
An important biomarker for predicting and diagnosing hypertension is the modified main  pulmonary arteries(PA), which is typically found in patients with pulmonary hypertension (PH) and has a diameter that is significantly bigger than that of a normal person. A broadening of the PA in chronic obstructive pulmonary disease (COPD) is linked to higher exacerbation risk and lower survival rates. Blockage of one of the pulmonary arteries, mostly caused by blood clots, causes  Pulmonary Embolism (PE). Therefore the early diagnosis of these pulmonary diseases, assessing the risk, and planning treatment for the patients at the early stage is required. Recently, advances in cardiac imaging have been accepted as good tools to help clinicians with early diagnosis and advanced planning in surgery. For the evaluation of many Pulmonary Vascular Diseases(PVD), the morphological examination of the Pulmonary Artery (PA) is crucial. However, even for specialists, the diagnosis procedure frequently takes a long time due to the enormous size and complexity of various imaging modalities; the pulmonary arteries must therefore be correctly and effectively segmented out.
Angiography offers insights into the blood flow and conditions of the vascular tree. Three-dimensional volumetric angiography information can be obtained using magnetic resonance (MRA), ultrasound, or x-ray-based technologies like computed tomography (CT). Currently, it is standard procedure to evaluate coronary and pulmonary artery diseases with computed tomography angiography (CTA) as it provides high-resolution 3D imaging as non-invasiveness. 
Various undermentioned reasons account for the difficulty of the pulmonary artery segmentation:
\begin{itemize}
\item The boundaries between the artery and background are often highly fuzzy. Also, Inside the lungs, we have arteries, veins, and airways that look very similar to each other. So, Making models that only learn to segment Arteries and ignore others will be a challenge.
\item The tubular structure of the coronary artery is extremely complex: the cross-section area changes gradually along the artery, and there exist a large number of bifurcations.
\item The appearance and geometry of the Pulmonary artery may vary considerably from one patient to another.i.e.one such reason is the buildup of the plaque or calcification inside the artery wall may further cause the variability from one patient to another
\item Finally, the image acquisition process may further introduce inherent image noise and artifacts, making the segmentation even more challenging. Also, data scarcity in the field of medical imaging has always been a bit of a challenge.
\item Creating a model that is more explainable, consumes less memory, and requires less inference time so that it can be deployed in real-time clinical applications.

\end{itemize}

The focus of our work is the accurate segmentation of the coronary artery in 3D pulmonary computed tomography angiography (CCTA) volumes. The segmentation of pulmonary artery structures benefits the quantification of its morphological changes for diagnosis of pulmonary hypertension with the task of maximizing the dice coefficient.

\section{Related Work}
Machine learning classifiers are used in a number of artery segmentation techniques, either for voxelwise airway classification or to eliminate false positive airway candidates from a leaky baseline segmentation. The multiscale Gaussian derivatives, multiscale Hessian-based features, or image texture features with local binary patterns are the set of predefined image features used by these classifiers (KNN, AdaBoost, support vector machines, or random forest). Compared to earlier, solely intensity-based algorithms, these techniques can produce more accurate vessel tree predictions with fewer false positives. However, they may take a long time to apply because they depend so heavily on the image features that were used to train the classifier in the first place. The primary benefit of deep CNN methods over traditional learning-based approaches is the automatic extraction of feature extraction from data in an end-to-end optimized environment. There are very few papers that talk about artery segmentations, so we studied some papers which describe segmenting out vessels or airways from the lungs CT/CTA. \cite{deep_vessel} talks about using 2-D orthogonal cross-hair filters, which make use of 3D context information at a reduced computational burden over 23{\%} improvement in speed, lower memory footprint, lower network complexity which prevents overfitting and comparable accuracy (with a Cox-Wilcoxon paired sample significance test p-value of 0.07 when compared to full 3-D filters). They also generate a synthetic dataset using a computational angiogenesis model capable of generating vascular trees under physiological constraints on local network structure and topology and use these data for transfer learning. \cite{a_v_classification} uses an algorithm that follows three main steps: first, a scale-space particle segmentation to isolate vessels; then, a 3-D convolutional neural network (CNN) to obtain the first classification of vessels; finally, graph cuts optimization to refine the results. The proposed algorithm achieves an overall accuracy of 94{\%}, which is higher than the accuracy obtained using other CNN architectures and Random Forest (RF). Inspired by recent ideas to use tree-structured long short-term memory (LSTM) to model the underlying tree structures for NLP tasks, \cite{TCT_connect}proposes a novel tree-structured convolutional gated recurrent unit (ConvGRU) model to learn the anatomical structure of the coronary artery. As the convolutions are used for input-to-state as well as state-to-state transitions, the tree-structured ConvGRU model considers the local spatial correlations in the input data, thus more suitable for image analysis.
\cite{fourth} uses a  two steps process: the first involves extracting the lung parenchyma using the Unet++ algorithm, which can significantly lower the over-segmentation rate, and the second involves extracting the pulmonary veins in the lung parenchyma using nnUnet.Only the blood vessels within the lung parenchyma are segmented, which eliminates interference from extracellular tissues and increases segmentation accuracy when the "AND" operation is done on the original image and the lung parenchyma segmentation results.
In the paper \cite{pu}, the author came out with a novel strategy to automatically identify and differentiate pulmonary arteries and veins depicted on chest CT without iodinated contrast agents. Initially, they used CNN to extrapulmonary arteries and veins and then applied a computational differential geometry method to automatically identify the tubular-like structures in the lungs with high densities, which are the intrapulmonary vessels. When compared to the results of a human expert, the computer algorithm labeled the pulmonary artery and vein branches with a sensitivity of almost 98 percent, proving the viability of the algorithm's automated system.
Lastly, the paper \cite{unet++} demonstrates a 2.5D seg-mentation network applied from three orthogonal axes, which presents a robust and fully automated pulmonary vessel segmentation result with lower network complexity and memory usage compared to 3D networks. The introduction of the slice radius convolves the central slice's nearby information, and multi-planar fusion enhances the presentation of intra- and inter-slice features by achieving a DICE score of 0.9272 and a precision of 0.9310. Inspired by all these papers we proposed our approach.

\section{Data Description}
The dataset is provided by PARSE 2022 Grand Challenge\cite{Parse-dataset}. The dataset contains 200 3D volumes provided in compressed NIFTI (.nii.gz) format and with refined pulmonary labels. These CTPA (Contrast Enhanced CT Pulmonary Angiography) are obtained from Harbin Medical University, Harbin, China. 
CTPA is a medical diagnostic test that employs computed tomography (CT) angiography to obtain an image of the pulmonary arteries. During the test, dye (normally Iodine contrast) will be injected into a vein that travels to pulmonary arteries. This dye makes the arteries appear bright and white on the scan pictures.
The size of the CT volumes ranges from 512*512*228 to 512*512*376. Pixel sizes are between 0.50mm/pixel and 0.95mm/pixel. Slice thickness is 1mm/pixel. The annotations are : 0 is referred to as Background and 1 as Pulmonary artery in voxel-level segmentation.
Out of these 200 3D volumes, 100 volumes are provided as a Training Dataset which consists of an image and respective label volume. 30 volumes for validation cases and 70 volumes for test cases.

\section{Methodology}
To perform segmentation of Pulmonary Arteries, we experimented with different encoder-decoder deep neural network architectures. We boiled down these architectures to two class models that were finally used to make predictions in this challenge's validation and test phase. These two categories of models are 3D UNet with residual units\cite{unet_3D} and Swin U-Net Transformer(Swin UNETR)\cite{swin_1,swin_2}. Our final predictions are made using a weighted ensemble of six models, out of which three are based on 3D UNet with residual units, and the other three are Swin UNETR trained in a different fashion in respect to each other. Details of architecture and training methods will be described later part of this section. 

\subsection{Pre-processing and Data Augmentation}
The CTA images were transformed and pre-processed for optimal results before passing them into the model. The Hounsfield unit(HU) of all the CT scans is clipped between -1000 HU to 1000 HU, and then these values were scaled to the range of 0 to 1. After performing clipping and scaling operations, extra slices containing no information were removed.  Following this, the most important pre-processing step was to convert the CT scan volume into smaller 3D volume patches. As discussed in the data description section, the input(CT scans) size ranges from 512x512x228 to 512x512x376, which cannot be passed entirely to 3D convolution because of high computational cost. Hence each CT scan volume is divided into smaller 3D cubes of patches. For U-Net based model, we experimented with different patch sizes of 96x96x96, 128x128x128, and 160x160x160. For the model, Swin UNETR, we stuck to the size of 96x96x96 because of computational cost.

Furthermore, to increase the total number of data points and to increase the robustness of the trained model augmentation technique like random flipping of volume along a different axis, random rotation within -30 to 30 degrees, and random intensity shift with a small offset of 10HU were performed.

\subsection{Model Architecture}
In total, six models were used to make a prediction, out of which three were U-Net based, and three were based on Swin UNETR. The model architecture of the 3D U-Net model had 16, 32, 64, 128, and 256 channels in subsequent layers with a stride of 2x2x2. The number of residual units was equal to 2, and the output layer had two channels, one representing the background and the other representing the Pulmonary Arteries. While training different U-Net-based models, we experimented with different patch sizes, augmentation techniques, and learning rates. Swin UNETR, the input size was set to 96x96x96, with input channels and output channels equal to 1 and 2, respectively. The feature size was set to 48. Similar to U-Net, three selected Swin UNETR used for prediction were trained with different learning rate, loss, and augmentation techniques. 

\subsection{Training and Evaluation}
 The training set for the PARSE challenge 2022 had in total of 100 CTA scans. We created a local validation set consisting of 10 CTA scans to train the model. Hence model was trained on 90 CTA images and validated on the remaining images. The training was performed using the ADAM optimizer with the loss function set as the sum of cross-entropy loss and Dice Loss. Using this configuration, both U-Net and Swin UNETR based were trained, and three of each category were chosen. 
 Finally, the weighted sum of predictions from each model was performed in the evaluation step. The calculations of weights and final prediction are explained in equations 1, 2, and 3.

Let D be a 1X6 matrix with each entry as a dice score from the corresponding model. So \[D \in R^{6X1}\] where Di represents the dice score of model i on the local validation set. Then ensembling weights, W is defined as in equation 1. 

\begin{equation}
        W = (1/(\sum Di))*D
    \end{equation}

This W was later used to make a weighted segmentation prediction, as explained in equations 2 and 3.
Let Pfinal be the final prediction, and Pi represents the segmentation result from model i, then,

\begin{equation}
        Pfinal = f(\sum Wi*Pi)
    \end{equation}
where f(x) is a thresholding function performed on all matrix elements and is defined in equation 3. 
\begin{equation}
    f(x) = \left\{
    \begin{array}{ll}
          0 & x\leq 0.5 \\
          1 & 0.5 \leq x \\
    \end{array} 
    \right.
\end{equation}

\section{Results and Discussion}
Table 1 describes the different models used for final prediction and their corresponding Dice score on the local validation set.
\begin{table}[]
\caption{Dice score of selected models on local validation set.}\label{tab1}
\begin{tabular}{|l|l|l|l|l|}
\hline
S.No & Model Name      & Patch Size  & Description                                                                                                                        & \begin{tabular}[c]{@{}l@{}}Dice Score on \\ local validation set\\ (In percentage)\end{tabular} \\ \hline
1    & Unet\_1         & 128x128x128 & U-Net model with no augmentations.                                                                                                 & 84.30                                                                                           \\ \hline
2    & Unet\_2         & 160x160x160 & \begin{tabular}[c]{@{}l@{}}U-Net model with bigger patch \\ size but with no augmentations.\end{tabular}                           & 85.50                                                                                           \\ \hline
3    & Unet\_3         & 160x160x160 & \begin{tabular}[c]{@{}l@{}}U-Net model with bigger patch \\ size but with augmentations.\end{tabular}                              & 85.52                                                                                           \\ \hline
4    & Swin\_UnetTr\_1 & 96x96x96    & \begin{tabular}[c]{@{}l@{}}Swin UNETR based architecture \\ with augmentations,\\ Learning rate = 1e-4\end{tabular}                & 86.55                                                                                           \\ \hline
5    & Swin\_UnetTr\_2 & 96x96x96    & \begin{tabular}[c]{@{}l@{}}Swin UNETR based architecture \\ with augmentations, learning rate = 1e-5\end{tabular}                  & 86.75                                                                                           \\ \hline
6    & Swin\_UnetTr\_3 & 96x96x96    & \begin{tabular}[c]{@{}l@{}}Swin UNETR based architecture \\ with more rigorous augmentations,\\  learning rate = 1e-5\end{tabular} & 86.87                                                                                           \\ \hline
\end{tabular}
\end{table}

In our experiments, we found that Swin UNETR gave a better dice score than U-Net based model. However, this difference is very small, with 1 percent of the dice score. Upon making multiple submissions to the validation phase, we found out that the Swin UNETR-based model performed better in segmenting Arteries’ branches while the U-Net model performed better in detecting the main Artery. Hence, to capture the benefit of both models, we performed ensembling to capture the main artery and branches better. For the U-Net model, the size of the input patch cube was one of the critical hyperparameters. One of the plausible explanations for this is that a larger patch size helps in better context and improved training. To carefully detect the branches and model needs more context. Because of the high computational cost, we cannot increase the input patch size from 96x96x96 for the Swin UNETR model. We also experimented with Post-processing techniques like the largest Connected Component Analysis(CCA). However, CCA was not used in the final pipeline as we concluded from our experiments that after applying CCA to the final prediction, the average dice score of the main artery increases, while for branches, the average dice score decreases. But the PARSE challenge used the Multi-level Dice Similarity Coefficient score, which is the weighted sum of the average dice score of branches and main artery with more weight given to branches. So reducing branch dice score after applying CCA was hurtful to our raking on the leaderboard. Finally, Figure 1, shows a 2 dimensional(2D) plot of one of the local validation set example. These 2D plots are generated by summation of the slices in corresponding, axial, sagittal or coronal, planes

\begin{figure}
\centering
\includegraphics[scale = 0.4]{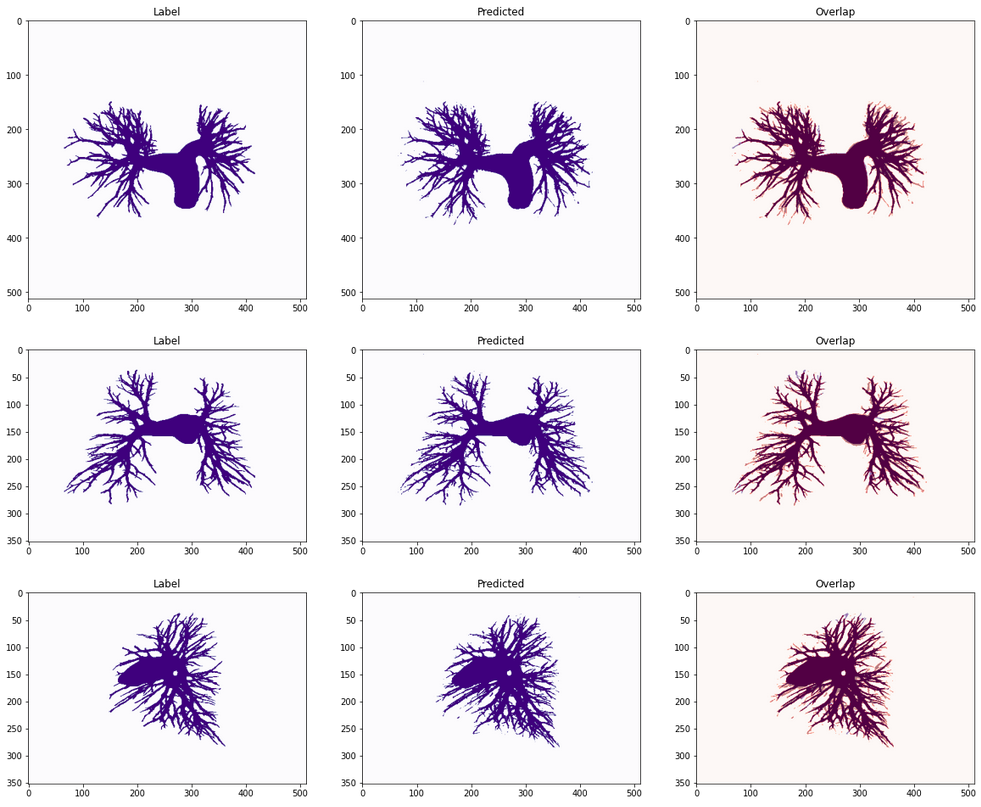}
\caption{Axial, Coronal and Sagittal 2 Dimensional plot of the label, ensemble model prediction and their overlap for one of the CT in local validation set. }
\end{figure}

\section{Acknowledgment}
We would like to thank Dr. Vinayak Rengan for his valuable input that helped us to understand the CTA scans better. We would also like to acknowledge Robert Bosch Center for Data Science and Artificial Intelligence (RBCDSAI), Indian Institute of Technology Madras, India, Project no: CR1718CSE001RBEIBRAV, for funding our project and also providing us with computational facilities.


\end{document}